\documentstyle[12pt,aaspp,flushrt]{article}
\newcommand{\Ref}{\hangindent=20pt \hangafter=1 \noindent}
\newcommand{\StartRef}{\hyphenpenalty=10000 \raggedright}
\newcommand{\rb}[1]{\raisebox{1.5ex}[0pt]{#1}}
\newcommand{\beq}{\begin{equation}}
\newcommand{\eeq}{\end{equation}}

\newcommand{\NarrowMargins}{
  \setlength{\oddsidemargin}{+0.3in}
  \setlength{\evensidemargin}{-0.0in}
  \setlength{\textwidth}{6.2in}
  \setlength{\topmargin}{-0.75in}
  \setlength{\textheight}{9.25in}   }
\catcode`\@=11 % This allows us to modify PLAIN macros.
\def\lsim{\mathrel{\mathpalette\@versim<}}
\def\gsim{\mathrel{\mathpalette\@versim>}}
\def\@versim#1#2{\vcenter{\offinterlineskip
        \ialign{$\m@th#1\hfil##\hfil$\crcr#2\crcr\sim\crcr } }}
\catcode`\@=12 % at signs are no longer letters
\NarrowMargins
\begin{document}
%\setlength{\baselineskip}{22pt}
%%%%%%%%%%%%%%%%%%%%%%%%%%%%%%%%%%%%%%%%%%%%%%%%%%%%%%%%%%%%%%%%%%%
%%%%%%%%%%%%%% ABSTRACT %%%%%%%%%%%%%%%%%%%%%%%%%%%%%%%%%%%%%%%%%%%  
%%%%%%%%%%%%%%%%%%%%%%%%%%%%%%%%%%%%%%%%%%%%%%%%%%%%%%%%%%%%%%%%%%%
\title{Gamma--ray Emission From Advection--Dominated Accretion Flows 
Around Black Holes: 
Application to the Galactic Center}
\author{Rohan Mahadevan, Ramesh Narayan}
\affil{Harvard-Smithsonian Center for Astrophysics,
60 Garden St., Cambridge, MA 02138 \and }
\author{Julian Krolik}
\affil{Department of Physics and Astronomy, Johns Hopkins University, Baltimore, MD 21218 }
\begin{abstract}
We calculate the flux and spectrum of $\gamma$-rays emitted by a
two-temperature advection-dominated accretion flow (ADAF) around a
black hole.  The $\gamma$-rays are from the decay of neutral pions
produced through proton-proton collisions.  We discuss both thermal
and power--law distributions of proton energies and show that the
$\gamma$-ray spectra in the two cases are very different.  We apply
the calculations to the $\gamma$-ray source, 2EG J1746-2852, detected
by EGRET from the direction of the Galactic Center.  We show that the 
flux and spectrum of this source are consistent with emission from an 
ADAF around the supermassive accreting black hole Sgr A$^*$ if the proton
distribution is a power--law.  The model 
uses accretion parameters within the range made likely by other 
considerations.
If this model is correct, it provides evidence for the presence of a 
two temperature plasma in Sgr A$^*$, and predicts $\gamma$--ray 
fluxes from other accreting black holes which could be observed with more 
sensitive detectors.
\end{abstract}
%%%%%%%%%%%%%%%%%%%%%%%%%%%%%%%%%%%%%%%%%%%%%%%%%%%%%%%%%%%%%%%%%%%
%%%%%%%%%%%%%% INTRODUCTION %%%%%%%%%%%%%%%%%%%%%%%%%%%%%%%%%%%%%%%  
%%%%%%%%%%%%%%%%%%%%%%%%%%%%%%%%%%%%%%%%%%%%%%%%%%%%%%%%%%%%%%%%%%%
\section{Introduction}
Advection--dominated accretion flows (ADAFs) are optically thin hot accretion
flows with low radiative efficiency 
(Narayan \& Yi 1994,
1995 a,b; Abramowicz et al. 1995).  
Unlike standard thin disks (see Frank et al. 1992) in which
the viscously generated 
energy is thermalized and radiated locally, ADAFs store most of the 
viscous energy and advect it into the central star. 
The gas in ADAFs  has  a two temperature structure 
(Shapiro, Lightman, \& Eardley 1976; Rees et al. 1982), 
with the ions being hotter than the
electrons.
The viscous heating affects  mainly the ions,
the more massive species, while the radiation is produced primarily by the electrons.
Since the ions transfer only a small fraction of
their energy to the electrons via
Coulomb scattering, the energy which is radiated is much less than 
the total energy released during accretion (Rees et al. 1982).

The emission spectrum of an ADAF
is mainly determined by the cooling processes of the electrons, 
viz. synchrotron, bremsstrahlung, and Compton processes
(see eg. Narayan \& Yi 1995b; Mahadevan 1997).  
Detailed calculations (e.g. Narayan 1996) show that
ADAFs around accreting black holes have a characteristic spectrum ranging from  
radio frequencies $\sim 10^{10} $ Hz up to hard X--ray frequencies $\sim 10^{21}$ Hz, 
whose shape is a function primarily of the mass accretion rate, and to some extent the mass of the accreting star.  A number of accreting black 
hole systems have been shown to contain ADAFs (e.g. Narayan, Yi, \& Mahadevan 1995; Narayan,
McClintock, \& Yi 1996; 
Lasota et al. 1996, Fabian \& Rees 1995, 
Mahadevan 1997, Narayan, Barret, \& McClintock 1997,
Reynolds et al. 1997).  

In the work done so far, on ADAFs, only the cooling of the electrons has
been considered in calculating the spectra.
However, since ADAFs are two temperature plasmas with
very high ion temperatures, e.g. $T_i \sim 10^{12}$ K close to a black hole 
(Narayan \& Yi 1995b),  one   wonders if there might be radiative processes 
associated directly with the ions. In particular, 
at such high temperatures,
collisions between protons can  lead to substantial  
production of neutral pions, $\pi^0$,  which would
decay into high energy $\sim 70 $MeV $\gamma$--rays.
Charged pions can also be produced by proton-proton collisions, but the
final decay products are neutrinos and electrons (and their anti-particles),
so that most of the
energy leaves the plasma either in neutrinos or lower energy photons.

Gamma--ray emission from a two temperature gas accreting onto 
a black hole has been computed previously (see  e.g. 
Dahlbacka, Chapline, \& Weaver 1974; 
Colpi, Maraschi, \& Treves 1986, Berezinsky \& Dokuchaev 1990). Most of the
previous work considered a thermal
distribution of protons, and the predicted
$\gamma$--ray fluxes
depend sensitively on the particular accretion scenario being considered.
In this paper we consider both thermal and non--thermal distributions of 
protons and focus specifically on the ADAF paradigm.
Using the methods described by Dermer (1986ab) we calculate the
flux and spectrum of $\gamma$--ray emission from ADAFs and show that these flows
produce interesting levels of high energy $\gamma$--rays. In the case of the
source Sagittarius A$^*$ (Sgr A$^*$) at the Galactic Center,
we predict a $\gamma$--ray flux which is above the sensitivity limit of 
the EGRET detector on the Compton Gamma--Ray Observatory,
and we compare the prediction with a possible detection of this source.

The outline of the paper is as follows.  In \S 2 we present the reaction rate 
for pion production and  describe 
the structure and basic equations of ADAFs.
In \S 3  we calculate the $\gamma$--ray spectra from  thermal and 
non--thermal distributions of protons in ADAFs, and in  \S 4 we apply these results to 
EGRET observations of the Galactic Center.
In \S 5 we discuss other applications of the results.
%%%%%%%%%%%%%%%%%%%%%%%%%%%%%%%%%%%%%%%%%%%%%%%%%%%%%%%%%%%%%%%%%%%
%%%%%%%%%%%%%% GENERAL RELATIONS %%%%%%%%%%%%%%%%%%%%%%%%%%%%%%%%%%  
%%%%%%%%%%%%%%%%%%%%%%%%%%%%%%%%%%%%%%%%%%%%%%%%%%%%%%%%%%%%%%%%%%%
\section{General Relations}
\subsection{Pion Production}
\label{genrels}
The production of $\gamma$--rays from $p-p$ collisions is a two 
step process, and has been worked out in detail by Stecker (1971, see also
Stephens \& Badhwar 1981; Dermer 1986ab). 
The colliding protons produce an intermediate particle, a 
neutral pion, $\pi^0$, which then decays into two $\gamma$--ray photons.  The reaction
is
\beq
p + p \rightarrow p + p  + \pi^0, \nonumber
\eeq
\beq
\pi^0 \rightarrow \gamma_1 + \gamma_2,
\eeq
where the photons, $\gamma_1, \ \gamma_2$, in general have different 
energies in the gas frame.  The number of neutral pions  produced is calculated by  
considering two protons with four momenta $p_1 $ and $p_2$, 
moving towards each other with relative 
velocity $v_{\rm{rel}}$.  
The number of collisions that occur in a volume $dV$, for a time $dt$, is
a frame invariant quantity,  which  in  an arbitrary reference frame 
can be written as (see Landau \& Lifshitz 1975, \S 12)
\begin{eqnarray}
dR_{1 \, 2} &=& \sigma \, v_{\rm{rel}} \, {p_1 \cdot p_2
        \over E_1 \, E_2} \, n_1 n_2 \, dV \, dt, \\
&=& c \sigma \, n_1 n_2 \, \sqrt{ (\vec{\beta_1} - \vec{\beta_2})^2 -
        (\vec{\beta_1} \times \vec{\beta_2})^2} \, dV \, dt.
\end{eqnarray}
Here, $n_1 $, $n_2$ are the number densities of the protons, 
$E_1,  \, E_2$ are their energies, 
$\sigma$ is the total cross--section for pion production, 
and $\vec{\beta_1}, \ \vec{\beta_2}$ are the
respective velocity parameters of the two particles.  For a distribution of 
particle velocities, the total number of pions
produced per second is just the integral of $dR_{1 \, 2}$ over volume and
velocities, which in spherical co--ordinates is 
\begin{eqnarray}
N(E_{\pi})   &=& 2\pi \, c  \, \int R^2 \, dR \, \int_1^{\gamma} \, d\gamma_1 \, 
\int_{\gamma}^{\infty}  d\gamma_2 \,  \int_{-1}^{1} \, d\cos\!\theta \nonumber \\
&\times& \sigma(\gamma_1, \gamma_2, \cos\!\theta) \,
n(R,\gamma_1) \, n(R,\gamma_2) \, \sqrt{ (\vec{\beta_1} -
\vec{\beta_2})^2 - (\vec{\beta_1} \times \vec{\beta_2})^2}
\ \ \mbox{s$^{-1}$},  \label{noproduction}
\end{eqnarray}
where $\gamma_1, \ \gamma_2$ are the respective Lorentz factors of the two
protons, $\cos\!\theta = \vec{\beta_1} \cdot \vec{\beta_2}/
|\vec{\beta_1}||\vec{\beta_2}|$, and $R$ is the radius with respect to the 
accreting star.
The total number of $\gamma$--rays produced is  $2 \, N(E_{\pi})$. 

To obtain the $\gamma$--ray spectrum, we first calculate the spectrum of
pion energies, $F(E_{\pi})\, dE_{\pi}$, in the frame of the observer.
  The observed  $\gamma$--ray spectrum is then obtained through the 
relation (Stecker 1971)
\beq
F(E_{\gamma}) = 2 \int_{E_{\pi \rm{min}}}^{\infty} dE_{\pi} { 
F(E_{\pi}) \over \sqrt{E_{\pi}^2 - m_{\pi}^2} }
\ \ \mbox{photons s$^{-1}$ GeV$^{-1}$},
\label{pitoga}
\eeq
where $m_{\pi} =  0.135$ GeV$/c^{2}$ is the mass of the pion, and 
$E_{\pi \rm{min}}$ is the minimum pion energy required to produce 
a $\gamma$--ray with energy $E_{\gamma}$,
\beq
E_{\pi \rm{min}} = E_{\gamma} + {m_{\pi}^2 \over 
		4 \, E_{\gamma}^2}.
\eeq

The pion spectrum, $F(E_{\pi})$ is obtained by 
substituting 
the differential cross--section 
$d\sigma (E_{\pi}; \gamma_1, \  \gamma_2, \cos\!\theta)/dE_{\pi}$ for 
$\sigma(\gamma_1, \gamma_2, \cos\!\theta)$ 
in equation (\ref{noproduction}). 
Two models are generally used to
determine the  differential cross--section  for pion production: 
the isobar model and the scaling model.  
In the isobar model, pions  are produced by a two step process. 
Colliding  protons produce an isobar, with rest mass 1.238 GeV, which
subsequently decays into
a proton and a pion (Lindenbaum \& Sternheimer 1957; Stecker 1971).  
Dermer (1986b) has shown that  the isobar model
agrees with the experimental data quite well for proton energies near 
threshold: $E_p \lsim 4$ GeV.  On the other hand,
for proton energies $E_p \gsim 8 $ GeV, the 
scaling model of Stephens and Badhwar (1981) represents the 
experimental data better (Dermer 1986b).
In our calculations, we therefore use the two models in their respective 
ranges of validity and interpolate smoothly 
(cubic interpolation) in the intermediate regime.

\subsection{ADAF Equations}

To evaluate equations (\ref{noproduction}) and (\ref{pitoga}) we need the
number density of protons as a function of energy in the accretion flow, $n_p(R,\gamma)$.  We write
$n_p(R,\gamma)$ as the product of a normalized velocity
distribution, which depends on the temperature at a given radius, and the 
total number density of protons,
\beq
n_p(R,\gamma) = n_p(R) \, n_{\gamma}[\gamma, \theta_p(R)], \label{3dreq8}
\eeq
with 
\beq
\int_{1}^{\infty}  n_{\gamma}[\gamma, \theta_p(R)] \, d\gamma = 1, \ \ \ \ \ 
\int_1^{\infty} (\gamma - 1) \, n_{\gamma}[\gamma, \theta_p(R)] \, d\gamma =  {3 \over 2} \, 
\theta_p(r).
\label{fdeq10}
\eeq
Here, $\theta_p(R)$ is the dimensionless ion temperature, $\theta_p = kT_i/m_p c^2$,
at radius $R$.

An ADAF is characterized by four parameters: the 
viscosity parameter, $\alpha$ (Shakura \& Sunyaev 1973), 
the ratio of gas pressure to  total pressure, $\beta_{\rm{adv}}$,
the mass of the central black hole, $M$, and the mass accretion rate, $\dot{M}$.
Given these parameters, the global structure and dynamics of ADAFs can be
calculated (Narayan,  Kato, \& Honma 1997; Chen, Abramowicz, \& Lasota 1997). 
Generally, ADAFs appear to have large
values of $\alpha \gsim 0.1$ (Narayan 1996), and the uncertainty in the 
value is a factor of a few. Many published ADAF models in the literature 
use $\alpha = 0.3$ (e.g. Narayan et al. 1996, 1997). If there is  
equipartition between 
magnetic and gas pressure, we expect  $\beta_{\rm adv} = 0.5$; 
this is the value we favor. However,  
for completeness we also 
consider the more extreme case $\beta_{\rm adv}$ = 0.95, 
where the magnetic pressure contributes negligibly ($\sim  1/20$) 
to  the total pressure. 

The radial structure of ADAFs  is well approximated by 
a series of concentric spherical shells, 
with the properties of the gas varying as a function of radius (Narayan \& Yi 1995a). 
The continuity equation gives 
$\dot{M} = 4\pi \, R^2 \, v(R) \,  \rho(R)$, where $v(R)$  is the radial velocity
of the gas, and $\rho(R)$ is the mass density.   Writing the 
radius in units of the Schwarzschild radius, $r = R/R_S$,
where $R_S = 2.95 \times 10^{5} \ m$ cm and $m = M/M_{\odot}$ is the 
mass of the central star in solar mass units, and 
assuming that the mass fraction of hydrogen is $X = 0.75$, 
we have 
\begin{eqnarray}
n_p(r) &=& {X \, \rho(r) \over m_p} = 1.9 \times 10^{19} \, m^{-1} \dot{m} \, r^{-2} \, 
\left[ { v(r) \over c} \right]^{-1},  \nonumber \\
&\equiv& m^{-1} \dot{m} \, \tilde{n}(r) \ \ \mbox{cm$^{-3}$}. \label{neeq}
\end{eqnarray}
Here $\dot{m} = \dot{M}/\dot{M}_{\rm{Edd}}$
is the accretion rate in Eddington units, with 
$ \dot{M}_{\rm{Edd}} = 1.39 \times 10^{18} \, m$ g s$^{-1}$, and 
$\tilde{n}(r)$ is defined so as to 
scale out the dependence of $n_p(r)$ on $m $ and $\dot{m}$.
Given $\alpha$, $\beta_{\rm adv}$, $m$ and $\dot{m}$, equation (\ref{neeq}) allows
us to calculate $n_p(r)$ provided we have $v(r)/c$.  We obtain $v(r)/c$ from 
the global ADAF solution of Narayan et al. (1997). 

The ion temperature $T_i(r)$ is obtained from  the total gas pressure $p_g$ 
(Narayan \&  Yi 1995b),
\beq
p_g = \beta_{\rm{adv}} \rho(r) \, c^2_s(r) = {\rho(r) k \, T_i \over 
		\mu_i \, m_p} + {\rho(r) k \, T_e \over \mu_e \, m_p},
\label{tempeq}
\eeq
where $c_s(r)$ is the isothermal  sound speed, and $\mu_i = 1.23, \ \mu_e = 1.14$ 
are the effective molecular weights of the ions and electrons respectively.
Since $T_i \gg T_e$ we 
neglect the second term to obtain
\beq
T_i(r) = 1.34 \times 10^{13} \, \beta_{\rm{adv}} \, \left[ 
		{c_s(r) \over c} \right]^2 \ \ \mbox{K}, \label{teeq}
\eeq
where $c$ is the velocity of light.  We obtain  $c_s(r)/c$ again 
from the global solution of the ADAF (Narayan et al. 1997).

Given $n_p(r) $ and $T_i(r)$ and an assumed functional form for the 
proton energy distribution $n_{\gamma}(\gamma, \theta_p)$, we can substitute  equation 
(\ref{3dreq8}) in equations (\ref{noproduction}) and (\ref{pitoga}) to calculate 
the $\gamma$--ray spectrum.
%%%%%%%%%%%%%%%%%%%%%%%%%%%%%%%%%%%%%%%%%%%%%%%%%%%%%%%%%%%%%%%%%%%
%%%%%%%%%%%%%% RESULTS %%%%%%%%%%%%%%%%%%%%%%%%%%%%%%%%%%%%%%%%%%%%  
%%%%%%%%%%%%%%%%%%%%%%%%%%%%%%%%%%%%%%%%%%%%%%%%%%%%%%%%%%%%%%%%%%%
\section{Results}
At present our understanding of viscous  heating is poor, and  
we do not know whether the process leads to  
a thermal or power--law distribution of proton energies, or perhaps some combination
of the two.  In this section we calculate the pion spectra by considering two 
different proton energy distributions: a relativistic Maxwell--Boltzmann
distribution and a  power--law distribution.  

\subsection{Thermal Distribution}

The proton temperature in an ADAF close to a black hole is marginally 
relativistic, $\theta_p = k \, T_p/m_p \, c^2 \lsim 0.2$.
At such temperatures, pions are produced primarily by protons 
in the tail of the Maxwell--Boltzmann  distribution, since only these particles
 have energies above the threshold needed  for pion production. 
Thus, the pion production is very sensitive to the proton temperature,
and since $\theta_p$ 
scales approximately as  $1/r$ (Narayan \& Yi 1995b; Mahadevan 1997), 
most of the production
is from the innermost radii of the ADAF,  $r \lsim 10$.
  
The protons at these temperatures have momenta near the  threshold for pion 
production, and do not exceed $\sim $ 1 GeV/$c$.
At such energies the isobar model (Lindenbaum \& Sternheimer 1957;
Stecker 1971; 
Dermer 1986a) is more accurate and we have therefore used it
to calculate the cross--sections and energy spectra.
In this model, the isobars  move along the initial directions of the colliding
protons in the center of momentum (CM) system. It is therefore convenient to 
transform variables in equation (\ref{noproduction}) via 
a Lorentz boost to the CM reference frame, followed
by a rotation so that the colliding protons are moving along 
the z--axis (Dermer 1984, 1986a).
Using equation (\ref{noproduction}) with equation (\ref{neeq}), and rewriting in
terms of the new variables 
(see Dermer 1984, 1986a for details), we obtain
\begin{eqnarray}
F(E_{\pi}) &=& { (2.95\times 10^5)^3 \, m \, \dot{m}^2 \, \pi \,  c \over m_{\pi} }
\  \  \int_1^{r_{\rm max}} \, dr \  {r^2 \, \tilde{n}^2(r)   \over 
\theta_p(r) \, K^2_2[1/\theta_p(r)]} \ 
\int_1^{\infty} \, d\gamma_r 
\, {(\gamma_r^2 - 1) \over [2(\gamma_r + 1) ]^{1/2}} 
\nonumber \\
&\times& 
\int_{1}^{\zeta/m_{\pi}} \, {d\gamma^{*}  \over \gamma^* \beta^* } \,  
{ d\sigma^*(\gamma^*; \gamma_r) \over d\gamma^*}   \ \ 
\{\exp[-q\gamma \gamma^*(1-\beta\beta^*)] - \exp[-q\gamma \gamma^*(1+\beta\beta^*)]\}, 
\nonumber \\
& & \ \ \ \ \mbox{photons s$^{-1}$ GeV$^{-1}$}.
\label{fdeq14}
\end{eqnarray}
Here $E_{\pi} = m_{\pi} \gamma$ is the pion energy in the observer's frame, 
$\gamma_r = \gamma_1\gamma_2(1-\beta_1 \beta_2 \cos\!\theta) $ is the relative
Lorentz factor of the two colliding protons, 
$q\equiv [2(\gamma_r+1)]^{1/2}/\theta_p(r)$, $ d\sigma^*(\gamma^*; \gamma_r)/
d\gamma^*$ is the differential cross--section for the production of a pion with 
Lorentz factor $\gamma^* $ in the CM frame  for a given $\gamma_r$, 
$\zeta \equiv (S - 4m_p^2 + m_{\pi}^2)/S^{1/2}$ where $S = 2m_p^2(\gamma_r +1)$ 
characterizes the strength of the collision, 
$K_2(1/\theta_p) $ is the modified Bessel function of order 2, and we have used
a Maxwell--Boltzmann energy  distribution for the protons. We take  $r_{\rm max} $ 
to be $10^3$; however, since the proton temperature decreases steeply with 
radius ($T_i \sim 1/r$)
the contribution to the pion spectrum from the gas at $r > 30$ is
negligible.

The calculation in equation (\ref{fdeq14}) gives the pion spectrum, $F(E_{\pi})$. 
Using equation (\ref{pitoga}),  we then obtain the 
$\gamma$--ray spectrum.
The results are shown in Fig.  1.  The 
plot corresponds to two values of the viscosity parameter $\alpha$, 0.1
and 0.3, and  two  values of $\beta_{\rm adv}$, 0.5 and 0.95. 
For a given $\alpha$, increasing $\beta_{\rm adv}$ leads to increasing 
gas pressure, and therefore to  increasing ion temperature (see eq. \ref{teeq}). 
Changing $\beta_{\rm adv}$ from 0.5 to 0.95 causes the temperature to go up 
by a factor $\sim  2$, and since the pion production is 
extremely sensitive to temperature, the $\gamma$--ray luminosity increases 
 by nearly two orders in magnitude.
Changes in $\alpha$ also affect the $\gamma$--ray luminosity, though less 
sensitively.  Since the total luminosity is 
proportional to $n_p^2$ which varies roughly as 
$\alpha^{-2}$ (Narayan \& Yi 1995b),  
we might expect the luminosity to increase with decreasing $\alpha$.  This effect 
is counterbalanced however by small changes in the temperature, 
and the 
luminosity  in fact decreases  slightly when $\alpha$ goes down from $0.3$ to $0.1$. 

\subsection{Non--Thermal Distribution}

In this section we 
consider a power--law distribution of proton energies described by a 
spectral index $s$.  Equation 
(\ref{neeq}) continues to hold, but $n[\gamma,\theta_p(r)]$ is no longer a Maxwell--Boltzmann
distribution.  To determine the form of $n[\gamma,\theta_p(r)]$, two 
requirements have to be satisfied (see eq. \ref{fdeq10}), namely 
the normalization condition
and  the requirement that the average  kinetic energy of the power--law distribution of 
protons, at each radius, be equal
to the average energy  of the protons as given 
by the local ion temperature. 

We model the energy distribution of protons as
\beq
n[\gamma, \theta_p(r)] \, d\gamma = \left\{ [1-\zeta(r)] \, \delta(\gamma - 1) +
(s-1) \, \zeta(r) \, \gamma^{-s}\right\} \, d\gamma, \label{ntdef}
\eeq
where a fraction $1 - \zeta(r)$ of the protons
have  $\gamma \sim  1$, and a fraction $\zeta(r)$ are in a power--law
tail with 
index $s$.  The distribution is properly normalized.  The fraction
$\zeta(r)$ is fixed by
the energy requirement:
\beq
\zeta(r)  = {3 \over 2 } \, (s-2) \,  \theta_p(r). \label{zetaeq}
\eeq
Since the exact energy distribution for the fraction $1-\zeta(r)$ of 
protons with energy below threshold is not important for the present
calculation, we have simplified the distribution 
to a $\delta$ function in equation (\ref{ntdef}).

To calculate  the total pion luminosity, $N(E_{\pi})$,
we substitute equation (\ref{ntdef}) in  equation (\ref{noproduction}).  
We then obtain three
different terms proportional to $(1 - \zeta)^2$, $(1-\zeta)\, \zeta$, and 
$\zeta^2$, respectively.
The $(1-\zeta)^2$ term gives no contribution since it  
corresponds  to protons with $\gamma \sim 1 $ 
colliding with one another.  These protons 
do not have enough energy to produce pions.  The $(1-\zeta) \, \zeta$ term 
corresponds to protons with $\gamma > 1$ colliding with protons with
$\gamma \sim 1$, and we denote the contribution by  $N_1(E_{\pi})$.
Similarly,  the $\zeta^2$ term corresponds to protons with $\gamma >1 $ 
colliding with protons with $\gamma >1$, and we denote it as $N_2(E_{\pi})$.   
We have 
\begin{eqnarray}
N_1(E_{\pi}) &=& 4 \pi c (2.95 \times 10^5)^3 \, m  \dot{m}^2   \, (s-1)
\int \, dr \,  r^2 \, \tilde{n}^2(r)
\, [1 - \zeta(r)] \, \zeta(r)  \nonumber \\
&\times&  
\int_{1}^{\infty} \sigma(\gamma)
\gamma^{-s} \, \beta \,
d\gamma, \nonumber \\
&\equiv& m \dot{m}^2 \ I(s, \alpha, \beta_{\rm adv}) \,  
\int_{1}^{\infty} \sigma_{\rm mb}(\gamma)
\gamma^{-s} \, \beta \, d\gamma\ \ \mbox{s$^{-1}$},
\end{eqnarray}
and
\begin{eqnarray}
N_2(E_{\pi}) &=& 4.84 \, m  \dot{m}^2  \, (s-1)^2 
\int \, dr \,  r^2 \, \tilde{n}^2(r) \, \zeta^2(r) \int_1^{\gamma} \, 
d\gamma_1 \,  \gamma_1^{-s}  \, 
\int_{\gamma}^{{\infty}} d\gamma_2 \, \gamma_2^{-s} 
\nonumber \\
&\times& 
\int_{-1}^1 \, d(\cos\!\theta) \  
\sigma_{\rm mb}(\gamma_1, \gamma_2, \cos\!\theta)  \,  \sqrt{ (\vec{\beta_1} -
\vec{\beta_2})^2 - (\vec{\beta_1} \times \vec{\beta_2})^2}
\ \ \mbox{s$^{-1}$},
\end{eqnarray}
where $\sigma_{\rm mb}$ is the cross--section in units of millibarns, and 
\beq
I(s, \alpha, \beta_{\rm adv}) \equiv 9.68 \, (s-1)  \, \int \, dr \,  r^2 
\, \tilde{n}^2(r) \, [1 - \zeta(r)]  \, \zeta(r). \label{sdr18}
\eeq
In an ADAF, we generally have $\zeta \ll 1$ so that  most 
of the protons have $\gamma \sim 1$ (cf.  eq. \ref{zetaeq}).  
We therefore expect most of the pion production and $\gamma$--ray luminosity
to come from $N_1(E_{\pi})$.
Table \ref{N1N2} gives numerical results and we see that $N_2(E_{\pi})$  is 
indeed negligible compared with $N_1(E_{\pi})$.

In calculating the pion spectrum, we consider only the contribution from 
$N_1(E_{\pi})$ and write
\begin{eqnarray}
F(E_{\pi}) &\simeq& {d N_1(E_{\pi}) \over d E_{\pi}} \nonumber \\
 &=& {I(s,\alpha, \beta_{\rm adv})  
\, m \, \dot{m}^2 \,  \over m_{\pi} } \, 
 \int_{1}^{{\infty}} \, d\gamma \,
\gamma^{-s} \, \beta \, {d\sigma_{\rm mb}(\gamma_{\pi}, \gamma) \over
                d \gamma_{\pi}}, \ \ \mbox{   photons s$^{-1}$ GeV$^{-1}$}.
\label{n1eq}
\end{eqnarray}
Unlike in the case of a thermal distribution 
where most of the protons have  energies either below or just above  
threshold,  in a 
power--law distribution the protons have a wide range of energy and some
protons 
are well above threshold.  We therefore use both the isobar and scaling models,
as described in \S \ref{genrels} 

Equation (\ref{n1eq}) reveals the dependence of the pion flux and energy
spectrum on the parameters: (1) The flux is proportional to $m \, \dot{m}^2$. (2)
The flux depends on $\alpha$, $\beta_{\rm adv}$ and $s$ through the function
$I(s, \alpha, \beta_{\rm adv})$ defined in equation (\ref{sdr18}). (3) The  
shape of the spectrum depends only on $s$ and is given by the integral in equation
(\ref{n1eq}).

Tables \ref{ibp5} and \ref{ibp95} give
$I(s,\alpha, \beta_{\rm adv})$ for different values of its 
parameters. For fixed $\alpha$, increasing $\beta_{\rm adv}$ leads to increasing
$I(s, \alpha, \beta_{\rm adv})$ since the temperature in the flow increases.
However, unlike the thermal case, the luminosity is not excessively
sensitive to 
$\beta_{\rm adv}$ since the number of protons above threshold is 
directly proportional to
$\theta_p$ (cf. eq. \ref{zetaeq}); therefore, a change in
$\beta_{\rm adv}$ by a 
factor of two changes the total luminosity only by $\sim 2$.
For fixed $\beta_{\rm adv}$ in a self-similar solution
(Narayan \& Yi 1994, 1995b), the density
varies in proportion to $\alpha^{-1}$, so the luminosity scales as
$\alpha^{-2}$.  However, in the more accurate
global solutions employed in this paper, the dependence of density---and
hence luminosity---on $\alpha$ is not strictly a power--law, and the overall
strength of the dependence is somewhat weaker than in the self-similar
solutions.  To illustrate this dependence, we list the numerical
values of $I(s,\alpha, \beta_{\rm adv})$ for 
three different choices of $\alpha$ in Tables \ref{ibp5} and \ref{ibp95}.

Figure 2 shows $\gamma$--ray spectra for different values of the
proton energy spectral index $s$ (we fix $\alpha = 0.3$, $\beta_{\rm adv} = 0.5$).  
The $\gamma$--ray spectral index at  high energies $E_{\gamma} \gsim 1$ GeV
is the same as the 
energy spectral  index of the protons $s$, as shown by Dermer (1986b).
For this reason, we investigated values of $s$ in the range 2.3 -- 3.3. 
The spectrum turns over at $E_{\gamma} \sim 70 $ MeV (approximately half the 
pion rest mass) and falls for lower 
photon energies. Comparing Figures 1 and  2, we
see that the total $\gamma$--ray luminosity
from a power--law energy 
distribution of protons is comparable to 
that obtained from a thermal distribution when  $\beta_{\rm adv} = 0.95$, 
but is very  much
higher for $\beta_{\rm adv} = 0.5$. In our view $\beta_{\rm adv}=0.5$ is the
natural choice since it corresponds to equipartition between gas and magnetic
pressure.
The $\gamma$--ray spectrum extends to much higher photon energies 
when the protons have a power--law distribution compared to the thermal case.
%%%%%%%%%%%%%%%%%%%%%%%%%%%%%%%%%%%%%%%%%%%%%%%%%%%%%%%%%%%%%%%%%%%
%%%%%%%%%%%%%% GALACTIC CENTER %%%%%%%%%%%%%%%%%%%%%%%%%%%%%%%%%%%%  
%%%%%%%%%%%%%%%%%%%%%%%%%%%%%%%%%%%%%%%%%%%%%%%%%%%%%%%%%%%%%%%%%%%
\section{Application to Sgr A$^*$}

The black hole candidate at the center of our Galaxy, Sagittarius
A$^*$ (Sgr A$^*$), appears to be a scaled down version of an AGN.  The
source is believed to consist of a black hole of mass
$M=2.45\times10^6M_\odot$ (Eckart \& Genzel 1996a, 1996b) accreting
gas from its surroundings.  

The mass accretion rate in Sgr A$^*$ has been estimated by various
methods (see Genzel et al. 1994 for a summary).  The accretion of wind
gas from neighboring stars (principally IRS 16) seems to be the most
likely scenario.  The accretion rate is determined by considering the
fraction of the wind that comes within the accretion radius of the
central black hole, but the estimate depends sensitively on the
assumed wind velocity.  Genzel et al. (1994) obtain an accretion rate
of $\dot{M} \simeq 6 \times 10^{-6}$ $M_{\odot}/$ yr$^{-1}$ which 
corresponds to $\dot{m} \sim 10^{-4}$ for $M = 2.45 \times 10^6 $ 
$M_{\odot}$, assuming a wind velocity of 1000 km s$^{-1}$ and mass
loss rate from winds of $\sim 3.5 \times 10^{-3}$ $M_{\odot}/$
yr$^{-1}$.  Melia (1992) uses a wind velocity of $600$ km s$^{-1}$ and
obtains an accretion rate of $\dot{M} \simeq 2 \times 10^{-4}$
$M_{\odot}/$ yr$^{-1}$ ($\dot{m} \sim 3\times 10^{-3}$).  The true
accretion rate probably lies between these two estimates, and we will
require our models to have $10^{-4}<\dot m<3\times10^{-3}$.  Note that
Lacy et al. (1982; see also Rees 1982)  
estimated a much higher mass accretion rate of $\dot
M\simeq 2\times10^{-3}$ $M_{\odot}/$ yr$^{-1}$ from stellar
disruptions at the Galactic Center, but argued that the accretion due
to this probably has a low duty cycle.

Sgr A$^*$ has an extremely low luminosity, $L\sim10^{37}~{\rm
erg\,s^{-1}}$, which corresponds to $\dot m\sim10^{-8}$ if the
accretion flow has a standard radiative efficiency of 10\%.  Such a
low $\dot m$ is in serious conflict with the estimate of $\dot m$
given in the previous paragraph.  Narayan, Yi \& Mahadevan (1995)
suggested that Sgr A$^*$ does accrete at a rate near the one estimated,
but in an advection-dominated mode.  Using a self-similar ADAF model,
they were able to
reconcile the low luminosity of the source with a relatively large
$\dot m$: $\sim 8 \times 10^{-4} \alpha$.  If $\alpha > 0.1$, the
accretion rate inferred on the basis of the ADAF model would be
in agreement with that estimated on the basis of gas supply.
Furthermore, they obtained a reasonable fit to the
observed spectrum from radio to hard X-ray frequencies.  In this
section, we calculate the $\gamma$--ray luminosity and spectrum of Sgr
A$^*$ due to pion production, and compare these predictions with
the flux seen by EGRET from the direction of the Galactic Center.

Among several unidentified sources detected by EGRET in the Galactic
plane (Merck et al. 1996), the source 2EG 1746-2852 is of particular
interest since it is spatially coincident with the Galactic Center.
There is no firm identification of this source, but it is point-like
to within the resolution of the instrument ($\sim 1^{\circ}$), it is 
$ \sim 10\sigma$ above the local diffuse emission,  and its spectrum 
differs significantly from the spectra of other unidentified EGRET sources.
Out of the 32 sources whose spectra are reported by Merck et al. (1996),
27 have spectral slopes $> 2.0$ (photon index), 
4 have spectral slopes $ \sim 1.9$ and one has a very hard 
spectrum with a spectral slope of 1.7.  
This last source is 2EG 1746--2852, and it is located exactly at 
the Galactic Center.  
Following Merck et al. (1996),
we make the reasonable assumption that  2EG 1746-2852 corresponds to 
 Sgr A$^*$.  Figure 3 shows the spectrum
of the source as measured by EGRET.  The dashed line is the best--fit
power--law obtained by Merck et al. (1996)
over the photon energy range 100--4000 MeV.  Their fitted spectral 
slope is $s = 1.7 \pm 0.1$.  At higher energies, the data suggest
a roll--over in the spectrum which would give a softer 
spectral index at higher energies $(E \gg 4$ GeV).

\subsection{Thermal Distribution}

Figure 3 compares the $\gamma$--ray spectrum from the thermal model with the EGRET data.
The  solid line corresponds to $\alpha = 0.3$, $\beta_{\rm adv} = 0.95$,
and the dotted line to $\alpha = 0.3$, $\beta_{\rm adv} = 0.5$.  
The black hole mass is taken to be 
$m = 2.45 \times 10^6$,  and the accretion rates in the two models have 
been varied such that the predicted spectra agree with  
the data point at $E  \sim 200 $MeV;   the accretion rates are indicated on 
the plot.

It is clear that the thermal model predicts a spectrum with
the wrong shape and 
fails to explain the EGRET detection at  
$E \sim  1$ GeV.  Moreover, for the preferred value of $\beta_{\rm adv} = 0.5$,
the model requires a rather high $\dot{m}$, which is much larger
than the  range
we consider reasonable (see above).
We conclude that the thermal model is inconsistent 
with the observations.

\subsection{Non--Thermal Distribution}

Fig. 4 compares  $\gamma$--ray spectra from various  power--law  models 
($\alpha = 0.3, \beta_{\rm adv} = 0.5$)
with the EGRET data.  
The different curves correspond to different values of the proton energy 
index $s$.  For each $s$, we have varied 
the accretion rate so as to obatin the closest agreement  with the observed spectrum.
The results ($5 \times 10^{-4} < \dot m < 9.2 \times 10^{-4}$), which are only weakly dependent on $s$, are given in Table \ref{arest}.
The accretion rates so inferred fall well within the range of accretion
rates estimated on the basis of the properties of nearby gas
($10^{-4} < \dot m < 3 \times 10^{-3}$, Genzel et al. 1994),
and are within factors of a few of the accretion rate estimated
on the basis of fitting
the radio to X-ray spectrum with the somewhat less accurate self-similar
version of the ADAF model ($\dot m \sim 8\times 10^{-4} \alpha$, 
Narayan, Yi, \& Mahadevan 1995).  

    Reproducing the power--law portion of the $\gamma$--ray spectrum
is a by--product of postulating a power--law proton energy distribution.  However,
it is striking in this context that the best--fit slope, $s \simeq 2.7$, is
identical to the slope of the low-energy cosmic ray distribution.  The models 
predict roll-overs at both the
high energy (few GeV) and low energy ($\sim 200$MeV) ends of
the spectrum.  At the high-energy end there is just a hint of such a
roll-over in the data, but at the low-energy end, the evidence seems somewhat
stronger.

	The similarity of the best--fit  spectral shape  to the 
diffuse Galactic $\gamma$--ray emission has two possible interpretations. 
First, since the diffuse emission is generally interpreted as being due
to cosmic ray protons striking thermal protons in the interstellar 
medium, a mechanism similar to what occurs in an ADAF, 
the EGRET detection of 2EG 1746--2852 could be interpreted as 
cosmic ray protons colliding with a compact dense cloud of gas that is
spatially  coincident with the Galactic 
Center.  This interpretation would require such a cloud to have a considerably
greater gas density (and possibly also cosmic ray density) than in
the interstellar medium only slightly farther away from the Galactic Center.
If virtually all the Galactic Center region $\gamma$-ray flux were due
to such a cloud, the power--law proton distribution function version
of the ADAF model would be put seriously in doubt.  Thermal proton
distribution function versions of the ADAF model (or entirely different
models) would not be seriously constrained if the
$\gamma$-rays coming from the Galactic Center prove not to have their
source in Sgr A$^*$.

     The alternative is that the 
observed $\gamma$--ray spectrum is in fact due to  Sgr A$^*$. In this case, 
not only is the deduced accretion rate roughly consistent with previous
estimates, but we also have the interesting result that whatever process
determines the cosmic proton 
energy distribution (shock acceleration ?) is also at work in ADAFs.  
We find the numerical coincidence between the prediction of the power--law
version of the ADAF model and the observed $\gamma$-ray flux striking
enough to justify pursuing this interpretation.

  To gauge just how strong the quantitative agreement is,
we must consider how much freedom the model is given
by adjustable parameters.  Three free parameters are significant.
One ($\dot m$) is fixed to within an order of magnitude
or so by other considerations (the amount of gas available to accrete,
and fitting the radio--X-ray spectrum).  Another ($\alpha$), while formally
unconstrained over a range of several orders of magnitude, is often
supposed to have a value quite near the one which we infer (i.e. between
0.1 and 1, cf. Narayan 1996).  Only the third ($s$) is chosen almost entirely on the
basis of
fitting the observed $\gamma$--ray spectrum.   Because the $\gamma$--ray
luminosity predicted by the ADAF model (assuming a power--law proton
energy distribution) approximately scales with ${\dot m}^2 \alpha^{-1.5}$ (the
actual scaling with $\alpha$ is only very roughly described by a power--law),
the combined
{\it a priori} uncertainty in $\dot m$ and $\alpha$ may 
be regarded as giving the $\gamma$--ray luminosity
predicted by the model a possible range of $\sim 100$.  
%%%%%%%%%%%%%%%%%%%%%%%%%%%%%%%%%%%%%%%%%%%%%%%%%%%%%%%%%%%%%%%%%%%
%%%%%%%%%%%%%% CONCLUSIONS %%%%%%%%%%%%%%%%%%%%%%%%%%%%%%%%%%%%  
%%%%%%%%%%%%%%%%%%%%%%%%%%%%%%%%%%%%%%%%%%%%%%%%%%%%%%%%%%%%%%%%%%%
\section{Discussion \& Conclusions}
Since it is not understood whether viscous heating produces a thermal
or power--law distribution of proton energies, we have calculated the
$\gamma$--ray emission of ADAFs corresponding to both types of
distributions.  We expect that the true energy distribution in any
given source will be bracketed by these two extremes.  Spectra corresponding
to intermediate models can be easily calculated by taking a weighted 
sum of the thermal and power--law models. 

If the proton distribution is thermal, the spectrum has a
characteristic shape with a peak at $E_{\gamma} \sim 70$ MeV and very
little emission at either lower or higher photon energies.  For the
kinds of proton temperatures expected in an ADAF, only a small
fraction of the protons (in the tail of the Maxwellian distribution)
have sufficient energy to produce pions.  Consequently, even a minor
change in the temperature, say by a factor of $\sim 2$, can modify the
$\gamma$--ray luminosity by orders of magnitude (cf. Figure 1).  Since
the flux is very sensitive to the gas temperature, a detailed
understanding of the physics of the ADAF in the region $r\lsim 10 $
(where most of the $\gamma$--ray luminosity originates) is necessary
in order to make testable predictions.  The present study does not
have the necessary accuracy for this.  General relativistic effects
become important at these radii and need to be included consistently.
In principle, if these effects are included, the $\gamma$--ray
spectrum could be used as a sensitive probe of the proton temperature.
One might even hope to distinguish between rotating and non--rotating
black holes, since the physical properties of the flow at $r \lsim 10$
will be different for the two cases.

If the protons have a power--law distribution of energies, the
$\gamma$--ray spectrum again peaks at $\sim 70$ MeV, but there is
significant emission at higher photon energies.  Indeed, the spectrum
asymptotically has a spectral slope $s$ which is equal to the
power--law index $s$ of the proton distribution (Dermer 1986b).  Thus,
the detection of a power--law $\gamma$--ray spectrum not only indicates
the nonthermal nature of the protons but also helps determine the
energy index.

In contrast to the thermal case, only half 
the $\gamma$--ray luminosity due to a
power--law distribution of proton energies originates 
from $r \lsim 10$, while the other half is emitted from $r \gsim 10$.
Further, the luminosity is not very sensitive to
changes in the temperature.  Therefore, our present understanding of
the physics of ADAFs is probably adequate for a
reasonable estimate of the $\gamma$--ray flux, and we can thus test
this version of the ADAF paradigm usefully against observations.

The Galactic Center source, Sgr A$^*$ presents an excellent
opportunity to test the model, since there is a good case for the
presence of an ADAF in this source (Narayan et al. 1995) and the
predicted $\gamma$--ray flux is large enough to be detectable with
current instruments.  Indeed, EGRET has detected an unresolved  source in the
Galactic Center region, 2EG J1746-2852 (Merck et al. 1996).  It is 
possible that this source corresponds to a dense compact gas cloud 
interacting with cosmic rays.  However, the source could equally well be 
associated with Sgr A$^*$.  The $\gamma$--ray flux of Sgr A$^*$ which
we calculate with our model using a power--law distribution of proton
energies is in good agreement with the observed flux of 2EG
J1746-2852  {\it given parameters which are almost determined on
independent grounds}.
In order to fit the flux, we need an Eddington-scaled
mass accretion rate of $\dot m\sim 5 $--$ 9 \times 10^{-4}$
(for $\alpha = 0.3$), close
to the estimate $\dot m\sim 8 \times 10^{-4}\alpha$
obtained by Narayan et al. (1995) from fitting
the lower energy spectrum due to the electrons.  It is also compatible with the
range of  $\dot m$ quoted by Genzel et al. (1994) on the basis of gas
motions in the vicinity of Sgr A$^*$: $10^{-4} < \dot m < 3\times 10^{-3}$.  
In addition, the pion-decay model for the Galactic Center $\gamma$-ray spectrum
predicts a roll over at energies below $\simeq 100$~MeV that may be
present in the observed spectrum. 

An exciting aspect of the present study is that it provides for the
first time a direct probe of the protons in hot accretion flows.  A
fundamental assumption in ADAF models is that the plasma is
two-temperature.  However, until now there has been  no direct test of this
assumption since all previous investigations of the emission from
ADAFs dealt only with the electrons.  The $\gamma$--ray emission we
have considered in this paper is due entirely to the protons.
Moreover, it requires that the protons have nearly virial energies in
order to be able to exceed the energy threshold for pion production.
The encouraging results we have obtained in the case of Sgr A$^*$
indicate that this source might have a two-temperature
plasma exactly as postulated in the models.  More detailed study of Sgr A$^*$
coupled with future detections of other sources (see below), could
strengthen the case for two-temperature accretion flows significantly.

Another exciting aspect of this study is that the $\gamma$--ray
spectrum from pion decay provides a direct window to the energy
distribution of the protons.  In the ADAF paradigm, viscous heat
energy goes almost entirely into the protons.  Furthermore, it is easy
to show that, at least at the low $\dot m$ expected in quiescent
systems, the protons have no energetically important
interactions either among themselves or with the electrons.
Therefore, each proton retains memory of all the heating events it has
undergone during the accretion, and so the energy distribution of the
protons directly reflects the heating processes present in the plasma.
In principle, with sufficiently sensitive observations, one could use
$\gamma$-ray measurements as a direct probe of viscous heating.  This
would be invaluable for the theory of hot accretion flows.

The ADAF model has been applied successfully to several other
low-luminosity black holes in addition to Sgr A$^*$.  The sources 
studied include
two X-ray binaries in quiescence, A0620--00 and V404 Cyg (Narayan, McClintock, \&
Yi 1996, Narayan, Barret \& McClintock 1997), and
several quiescent AGNs, viz. the LINER galaxy NGC 4258 (Lasota et al. 1996,
Herrnstein et al. 1996) and the nearby elliptical galaxies NGC 4472,
NGC 4486, NGC 4649, and NGC 4636 (Mahadevan 1997; Reynolds et
al. 1997).  Table \ref{prediction}  gives the values of $m$, $\dot m$,
$\alpha$ and $\beta_{\rm adv}$ of several of these systems (taken from the
references indicated) and presents the expected $\gamma$--ray fluxes
according to our model.  Among the nearby ellipticals, we have
included only NGC4486 since it is the only system with a reliable mass
estimate (Ford et al. 1995; Harms et al. 1995).  The $\gamma$-ray flux
estimates in Table \ref{prediction}  assume that the protons have a power--law energy
distribution with $s= 2.75$.  We see that, only in the case of Sgr A$^*$ does
the detection threshold of EGRET
($10^{-8}~{\rm photons\,cm^{-2}\,s^{-1}}$) permit a test of the
predictions, although some of the other sources
might be not far below the EGRET threshold.  These sources are potentially
detectable with future instruments such as the Gamma Ray Large Area
Space Telescope (GLAST), which is designed to be 100 times more
sensitive than EGRET.

   Narayan (1996) has argued that X-ray binaries in the ``hard state''
(also called the ``low state'') may also contain ADAFs with $\dot m\lsim0.1$.
If the proton energy distribution in these sources is a power--law of
the sort we suggest exists in Sgr A$^*$, several of them might be
expected to produce detectable $\gamma$-ray fluxes.  However, the
accretion rates in these objects (e.g. Cyg X-1) is likely to be close to  
the maximum rate permitting an ADAF (Narayan \& Yi 1995b).
If this is the
case, particle interactions might be rapid enough that maintenance of a
true power--law distribution function would be questionable.  Accurate
predictions of the $\gamma$-ray flux would then require a more elaborate
calculation than we have performed for Sgr A$^*$.
 
Finally, we note that in determining the $\gamma$--ray spectra, we
have neglected gravitational redshift effects which become important
at $r \lsim 5$.  We have also neglected the Doppler blueshift
associated with the large radial and orbital velocities of the gas at
these radii (Narayan \& Yi 1995b).  Only a detailed calculation can
tell which of the two effects predominates.  We note, however, that
these effects are not likely to modify the results we have presented
for a power--law distribution of protons because more than half the
emission in this case occurs at $r \gsim 10$.

\noindent{\it Acknowledgments.}  RM thanks Areez Mody for useful
discussions. This work was supported in part by NSF grant AST 9423209.
J.H.K.'s research was supported in part by NASA Grant NAGW-3156.
\clearpage
%%%%%%%%%%%%%%%%%%%%%%%%%%%%%%%%%%%%%%%%%%%%%%%%%%%%%%%%%%%%%%%%%%%%%%
%%%%%%%%%%%% REFERENCES      %%%%%%%%%%%%%%%%%%%%%%%%%%%%%%%%%%%%%%%%%
%%%%%%%%%%%%%%%%%%%%%%%%%%%%%%%%%%%%%%%%%%%%%%%%%%%%%%%%%%%%%%%%%%%%%%
\newpage
{
\footnotesize
\StartRef
\noindent {\large \bf References} \\
\Ref Abramowicz, M., Chen, X., Kato, S., Lasota, J. P, \& Regev, O., 1995,
ApJ, 438, L37 \\
\Ref Berezinsky, V. S., \& Dokuchaev, V. I., 1990, ApJ, 361, 492-496 \\
\Ref Chen, X., Abramowicz, M., \& Lasota, J. P., 1997, ApJ, submitted \\
\Ref Colpi, M., Maraschi, L., \& Treves, A., 1986, ApJ, 311, 150-155 \\
\Ref Dahlbacka, G. H., Chapline, G. F., \& Weaver, T. A., 1974, Nature, 250, 
37 \\
\Ref Dermer, C., 1984, ApJ, 280, 328--333 \\
\Ref Dermer, C., 1986a, ApJ, 307, 47--59 \\
\Ref Dermer, C., 1986b, A\&A, 157, 223--229 \\
\Ref Eckart, A., \& Genzel, R., 1996a, in `` The Galactic Center'', ASP 
Conference Series, Vol 102, R. Gredel (ed.), 196 \\
\Ref Eckart, A., \& Genzel, R., 1996b, MNRAS, in press \\
\Ref Fabian, A. C., \& Rees, M. J., 1995, MNRAS, 277, L55-L58 \\
\Ref Ford, H. C. et al., 1995, ApJ, 435, L27 \\
\Ref Frank, J., King, A., \& Raine, D., 1992, Accretion Power in
Astrophysics (Cambridge: Cambridge Univ. Press) \\
\Ref Genzel, R., Hollenbach, D., \& Townes, C. H., 1994, Rep. Prog. Phys., 57, 417 \\
\Ref Harms, R. J. et al., 1994, ApJ, 435, L35 \\
\Ref Lacy, J. H., Townes, C. H., \& Hollenbach, D. J., 1982, ApJ, 262, 120 \\
\Ref Landau, L. D., \& Lifshitz, E. M., 1975, The Classical Theory of Fields, 
4th Ed., Pergamon, Oxford \\
\Ref Lasota, J. P., Abramowicz, M. A., Chen, X., Krolik, J.,
                Narayan, R., \& Yi, I. 1996, ApJ, 462, 142\\
\Ref Lindenbaum, S. J., Sternheimer, R. M., 1957, Phys. Rev., 105, 1874 \\
\Ref Mahadevan, R., 1997, ApJ, 477 \\
\Ref Melia, F., 1992, ApJ, 387, L25  \\
\Ref Merck et al., 1996, A\&A, in press\\
\Ref Narayan, R., 1996, ApJ, 462, 136 \\
\Ref Narayan, R., Barret, D., \& McClintock, 1997, ApJ, submitted \\
\Ref Narayan, R., Kato, S., Honma, F., 1997, ApJ, submitted \\ 
\Ref Narayan, R., McClintock, J. E., \& Yi, I., 1996, ApJ, 457, 821-833 \\
\Ref Narayan, R., \& Yi, I., 1994, ApJ, 428, L13 \\
\Ref Narayan, R., \& Yi, I., 1995a, ApJ, 444, 231 \\
\Ref Narayan, R., \& Yi, I., 1995b, ApJ, 452, 710-735 \\
\Ref Narayan, R., Yi, I., \& Mahadevan, R., 1995, Nature, 374, 623-625 \\
\Ref Rees, M. J., 1982, Nature, 333, 523 \\
\Ref Rees, M. J., Begelman, M. C., Blandford, R. D., \& Phinney, E. S.,
1982, Nature, 295, 17 \\
\Ref Reynolds, C. S., Di Matteo, T., Fabian, A. C., Hwang, U., \& 
Canizares, C. R., 1997, MNRAS, in press \\
\Ref Shakura, N. I., \& Sunyaev, R. A., 1973, A\&A, 24, 337 \\
\Ref Shapiro, S. L., Lightman, A. P., \& Eardley, D. M. 1976, ApJ, 204,
 187 \\
\Ref Stecker, F. W., 1971, Cosmic Gamma Rays (Baltimore: Mono Book Co.) \\
\Ref Stephens, S. A., \&  Badhwar, G. D., 1981, Astrophysics and Space Sciences, 
76, 213 \\

}

\newpage
%%%%%%%%%%%%%%%%%%%%%%%%%%%%%%%%%%%%%%%%%%%%%%%%%%%%%%%%%%%%%%%%%%%
%%%%%%%%%%%%%% TABLES %%%%%%%%%%%%%%%%%%%%%%%%%%%%%%%%%%%%%%%%%%%  
%%%%%%%%%%%%%%%%%%%%%%%%%%%%%%%%%%%%%%%%%%%%%%%%%%%%%%%%%%%%%%%%%%%
\begin{table}[h]
\caption[integral]{Photon luminosities (${\rm photons\,s^{-1}}$) in units of
$m\dot m^2$ for $\alpha = 0.3$, $\beta_{\rm adv} = 0.5$, and different 
power--law indices $s$. } 
\begin{center}
\begin{tabular}{lccc}\hline \label{N1N2} 
& & &  \\[1.5ex]
\rb{$s $} & \rb{$\displaystyle{{N_1(E_{\gamma}) \over m \dot{m}^2}}$} & 
\rb{$\displaystyle{{N_2(E_{\gamma}) \over m \dot{m}^2}}$} &
\rb{$\displaystyle{{N(E_{\gamma}) \over m \dot{m}^2}}$} \\ \hline \hline
2.1 & 7.36e+39 & 1.52e+37 & 7.38e+39  \\ 
2.2 & 1.26e+40 & 5.27e+37 & 1.27e+40  \\ 
2.3 & 1.65e+40 & 1.04e+38 & 1.66e+40  \\ 
2.4 & 1.92e+40 & 1.62e+38 & 1.94e+40  \\ 
2.5 & 2.12e+40 & 2.23e+38 & 2.14e+40  \\ 
2.6 & 2.26e+40 & 2.86e+38 & 2.29e+40  \\ 
2.7 & 2.36e+40 & 3.48e+38 & 2.39e+40  \\ 
2.8 & 2.42e+40 & 4.11e+38 & 2.46e+40  \\ 
2.9 & 2.47e+40 & 4.61e+38 & 2.55e+40  \\ 
3.0 & 2.49e+40 & 5.23e+38 & 2.54e+40  \\ 
3.1 & 2.50e+40 & 5.72e+38 & 2.55e+40  \\ 
3.2 & 2.49e+40 & 6.15e+38 & 2.55e+40  \\ 
3.3 & 2.47e+40 & 6.80e+38 & 2.54e+40  \\ 
3.4 & 2.45e+40 & 7.01e+38 & 2.52e+40  \\ 
3.5 & 2.42e+40 & 7.50e+38 & 2.49e+40  \\  \hline
\end{tabular}
\end{center}
\end{table}

\begin{table}
\caption[integral]{Selected values of $I(s, \alpha, \beta_{\rm adv})/10^{40}$
for $\beta_{\rm adv}=0.5$} 
\begin{center}
\begin{tabular}{lcccc}\hline \label{ibp5} 
& & &    \\[1.5ex]
\rb{$\alpha $} & \rb{$s = 2.1$} & 
\rb{$s= 2.3$} &
\rb{$s= 2.75$}  & \rb{$s = 3.3$}\\ \hline \hline
0.03 & 0.46 & 1.6 & 5.4 & 12.  \\ 
0.1 & .11 & 0.40 & 1.3 & 3.0  \\ 
0.3 & 0.024 & 0.082 &  0.28 & 0.62 \\  \hline
\end{tabular}
\end{center}
\end{table}

\begin{table}
\caption[integral2]{Selected values of $I(s, \alpha, \beta_{\rm adv})/10^{40}$
for $\beta_{\rm adv}=0.95$} 
\begin{center}
\begin{tabular}{lcccc}\hline \label{ibp95} 
& & &    \\[1.5ex]
\rb{$\alpha $} & \rb{$s = 2.1$} & 
\rb{$s= 2.3$} &
\rb{$s= 2.75$}  & \rb{$s = 3.3$}\\ \hline \hline
0.03 & 0.88 & 3.2 & 10. & 24  \\ 
0.1 & .15 & .52 & 1.7 & 3.8  \\ 
0.3 & 0.036 & 0.13 &  0.42 & 0.94 \\  \hline
\end{tabular}
\end{center}
\end{table}

\begin{table}
\caption[integral3]{Mass accretion rates required in Sgr A$^*$ in order to fit
the observed $\gamma$--ray spectrum.  The models assume
$m=2.45\times10^6$, $\alpha = 0.3, \beta_{\rm adv} = 0.5$, and a power--law
distribution of protons with index $s$.} 
\begin{center}
\begin{tabular}{lcc}\hline \label{arest} 
 & &    \\[1.5ex]
\rb{$s $} & 
\rb{$\dot{m}$ ($m = 2.45 \times  10^6$)} &
\rb{$\dot{M} \ \left(M_{\odot}/ \mbox{yr} \right)$} \\ \hline \hline
2.1 & 9.2 $\times 10^{-4}$& 5.2 $\times 10^{-5}$  \\ 
2.3 & 6.6  $\times 10^{-4}$ & 3.5 $\times 10^{-5}$  \\ 
2.75 &  5.2 $\times 10^{-4}$ &  2.8 $\times 10^{-5}$ \\ 
3.3 & 4.7 $\times 10^{-4}$ &  2.5 $\times 10^{-5}$ \\  \hline
\end{tabular}
\end{center}
\end{table}

\begin{table}
\caption[integral3]{Flux of photons $> 100$ MeV from various accreting
black holes with ADAFs ($s$ = 2.75).} 
\begin{center}
\begin{tabular}{lllllll}\hline \label{prediction} 
& & & & & & \\[1.5ex]
\rb{Name} & $\alpha$ & $\beta_{\rm adv}$ & $m$ & $\dot{m}$ & $D_{\rm kpc}$ & Flux (photons 
cm$^{-2}$ s$^{-1}$)\\ \hline \hline
Sgr A$^*$& 0.3 & 0.5 & 2.45 $\times 10^6$ & 5.2 $\times 10^{-4}$ 
& 8.5 & 4.9 $\times 10^{-7}$ \\ \hline 
A0620--00 & 0.3 & 0.5 & 6 & 1.2$\times 10^{-3}$ & 1 & 2.8$\times 10^{-9}$ \\
V404Cyg & 0.3 & 0.5 & 12 &  4.6$\times 10^{-3}$ & 3  & 1.5$\times 10^{-9}$ \\
NGC 4486 & 0.3 & 0.5 & 3$\times 10^9$ & 10$^{-2.5}$ & 16$\times 10^3$ & 
6$\times 10^{-9}$ \\ 
NGC 4258 & 0.3 & 0.95 & 3.6$\times 10^{7}$ & 10$^{-2}$& 6.5$\times 10^3$ & 
1.4$\times 10^{-9}$ \\ \hline

\end{tabular}
\end{center}
\end{table}

\clearpage
%%%%%%%%%%%%%%%%%%%%%%%%%%%%%%%%%%%%%%%%%%%%%%%%%%%%%%%%%%%%%%%%%%%
%%%%%%%%%%%%%% FIGURE CAPTIONs %%%%%%%%%%%%%%%%%%%%%%%%%%%%%%%%%%%%  
%%%%%%%%%%%%%%%%%%%%%%%%%%%%%%%%%%%%%%%%%%%%%%%%%%%%%%%%%%%%%%%%%%%
\newpage
\noindent{\bf Figure Captions.} \\

\noindent Figure 1:  $\gamma$--ray spectra from  a thermal distribution of
colliding protons for two different values each of $\alpha$ and
$\beta_{\rm adv}$.

\noindent Figure 2: $\gamma$--ray spectra from  a power--law 
distribution of colliding protons for different values of the energy index
$s$.  The curves correspond to $\alpha = 0.3$, $\beta_{\rm adv}$ = 0.5. 
Tables \ref{ibp5} and \ref{ibp95} can be used to obtain fluxes
for other values of $\alpha$ and $\beta_{\rm adv}$.

\noindent Figure 3: $\gamma$--ray spectra from  a thermal distribution of 
colliding protons in the Galactic Center source Sgr A$^*$.  The data
correspond to EGRET observations of the unidentified source 2EG
J1746-2852, which has been identified with Sgr A$^*$ (Merck et
al. 1996).

\noindent Figure 4: Similar to Figure 3, but for a power--law 
distributions of proton energy.  The curves correspond to different
values of the proton spectral index $s = 2.1, 2.3$, and 2.75.
Note the significantly superior fit compared to Figure 3.
The mass accretion rates of the various models are listed in
Table 4.
\newpage
\pagestyle{empty}
\begin{figure}
\epsffile{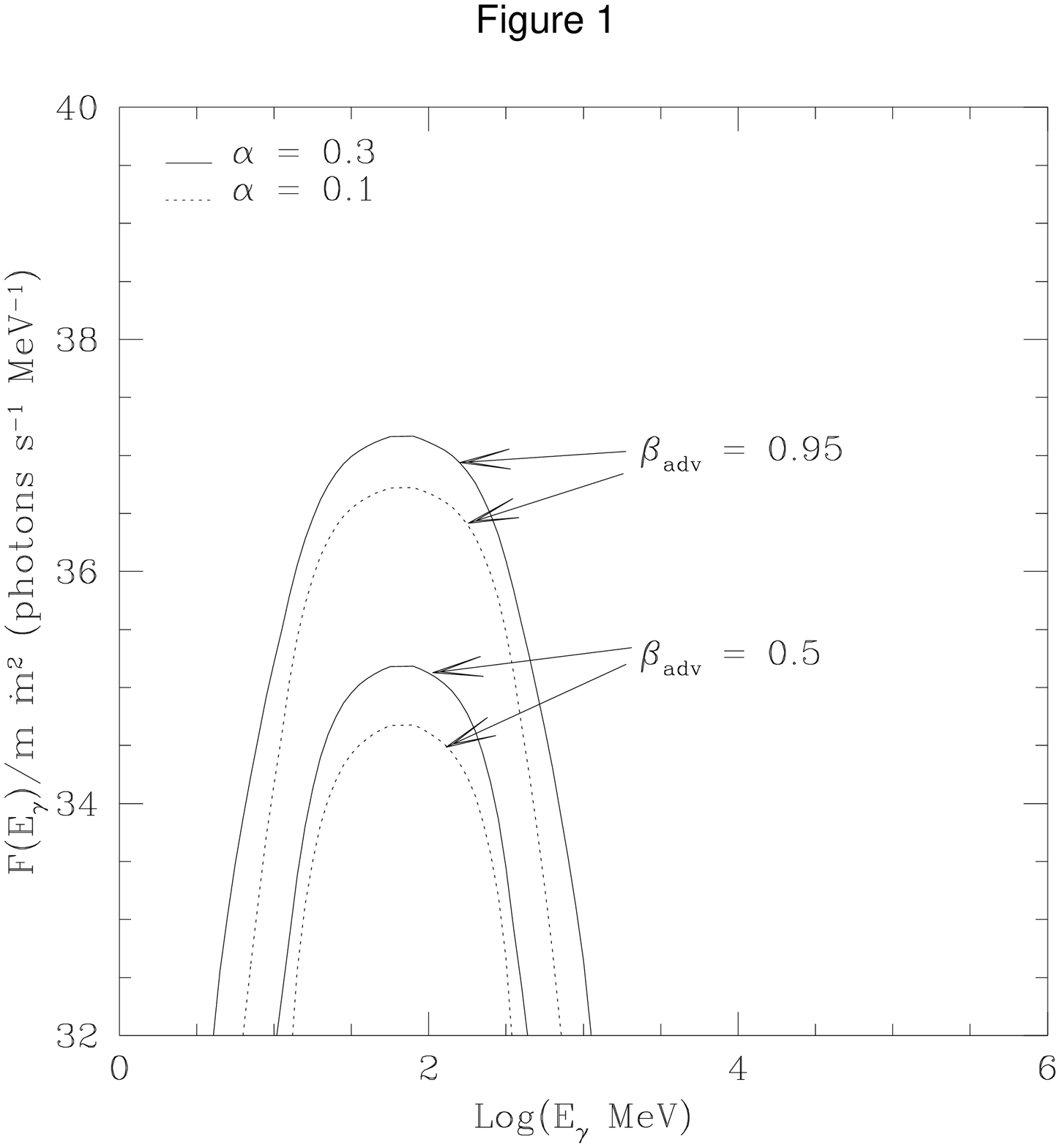}
\end{figure}
\begin{figure}
\epsffile{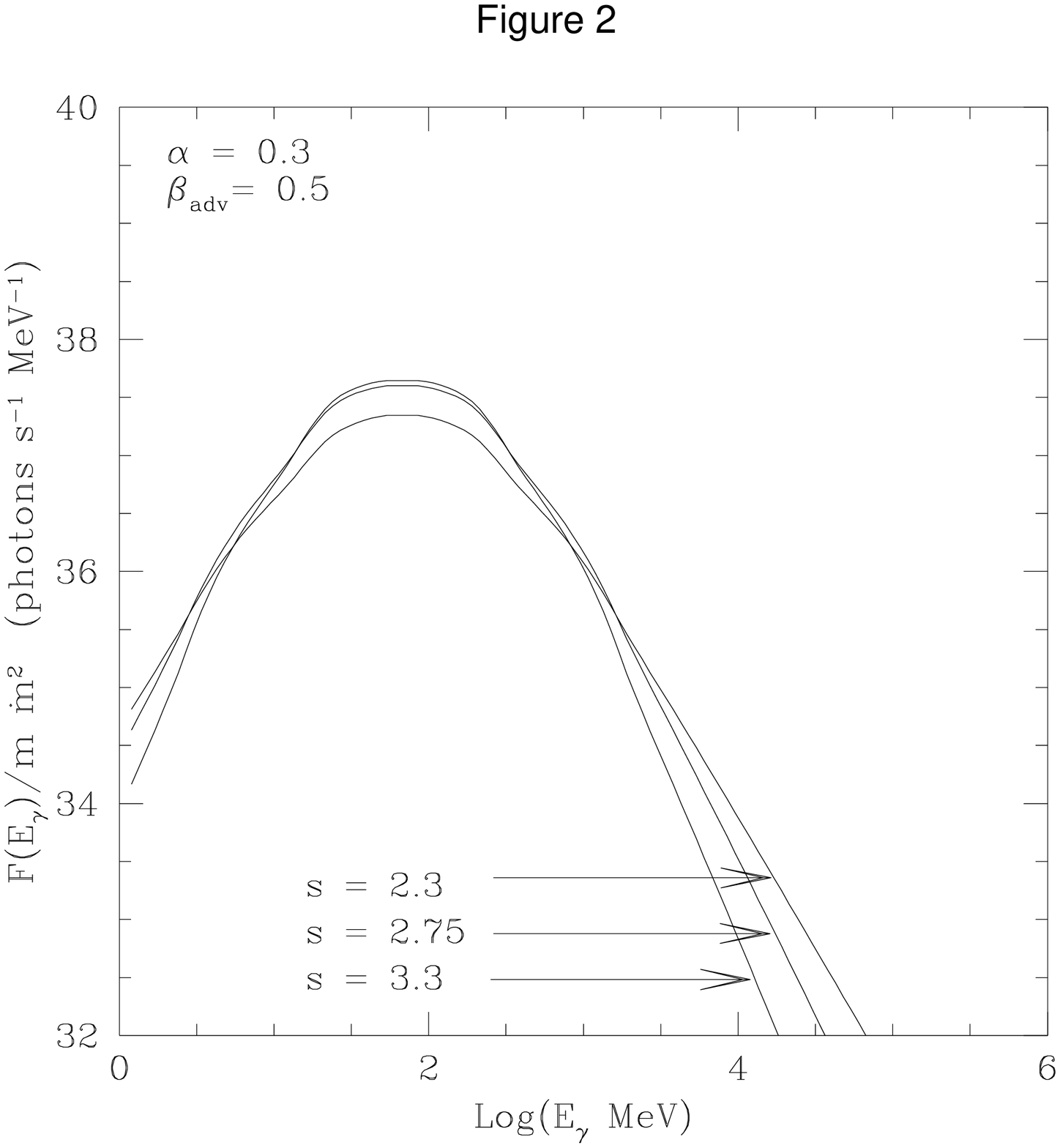}
\end{figure}
\begin{figure}
\epsffile{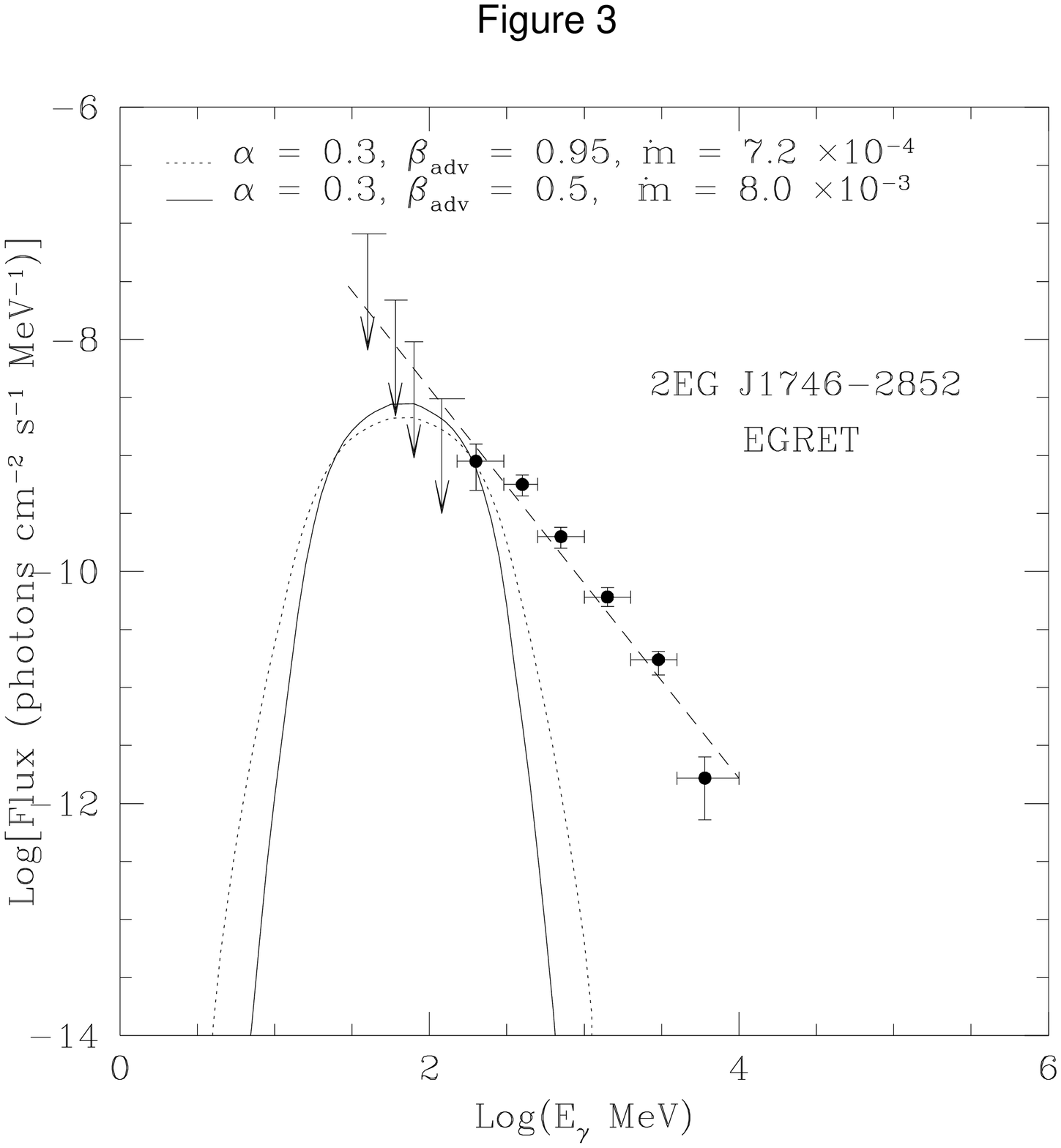}
\end{figure}
\begin{figure}
\epsffile{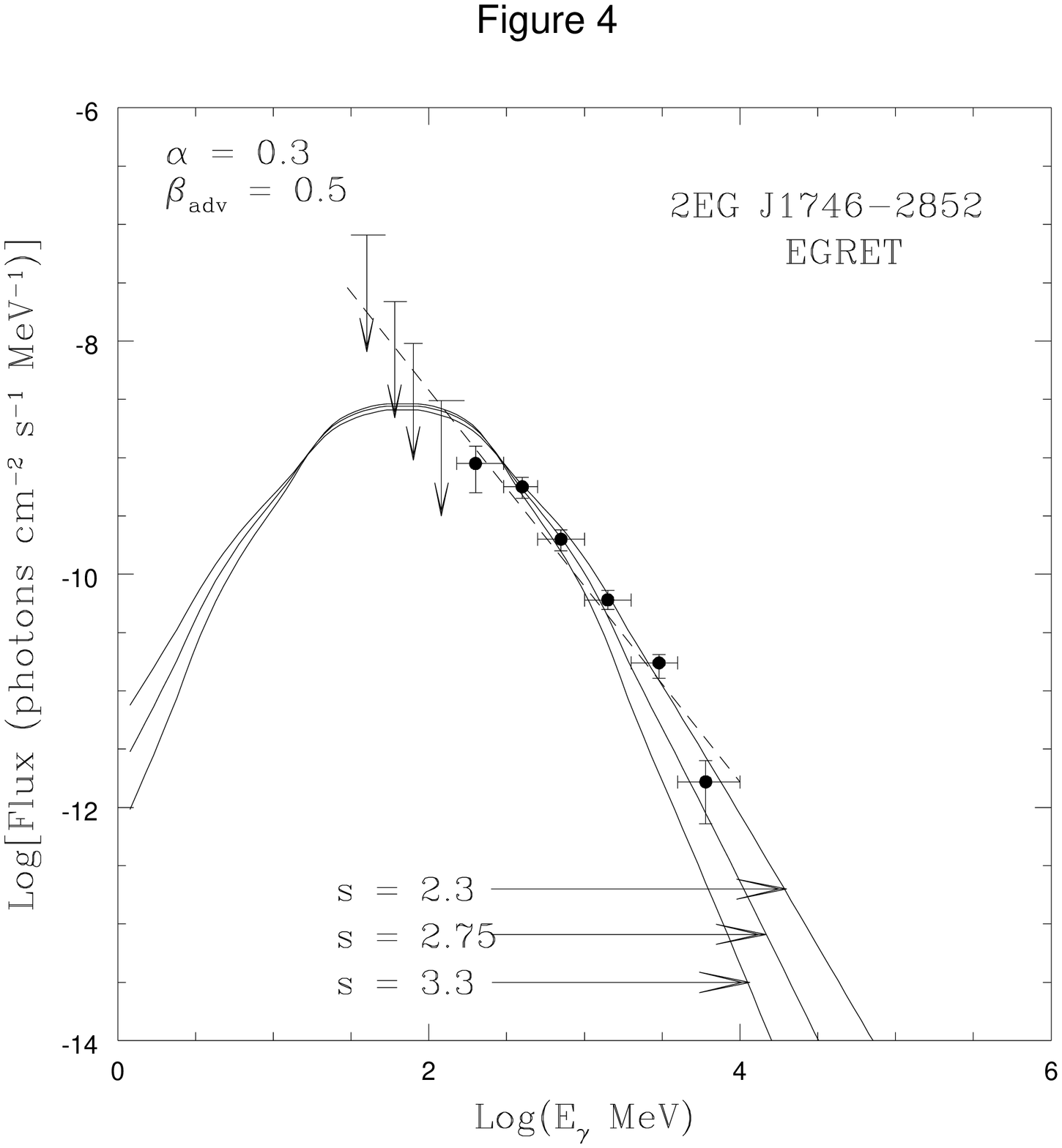}
\end{figure}
\end{document}